\documentclass{ws-rv9x6}
\usepackage{subfigure}     % required only when side-by-side / subfigures are used
\usepackage{ws-rv-van}     % numbered citation/references (default)

\usepackage{feynmp}
\usepackage{multirow}
\usepackage{bigstrut}
\usepackage{array}
\usepackage{fix-cm}

\newcommand{\nC}{\mathbf{C}}

\newcommand{\nG}{\mathbf{G}}

\newcommand{\vlowk}{V_{{\rm low}\,k}}

\makeindex
\begin{document}

\chapter*{Pairing in finite nuclei from low-momentum two- and three-nucleon interactions}\label{ra_ch1}

\author[T. Duguet]{T. Duguet}
%\index[aindx]{Author, F.} % or \aindx{Author, F.}
%\index[aindx]{Author, S.} % or \aindx{Author, S.}

\address{CEA-Saclay, IRFU/Service de Physique Nucl\'eaire, F-91191 Gif-sur-Yvette, France, \\
National Superconducting Cyclotron Laboratory
  and Department of Physics and Astronomy, 
  Michigan State University, East Lansing, MI 48824, USA, \\
thomas.duguet@cea.fr}

\begin{abstract}
The present contribution reviews recent advances made toward a microscopic understanding of superfluidity in nuclei using many-body methods based on the BCS ansatz and low-momentum inter-nucleon interactions, themselves based on chiral effective field theory and renormalization group techniques. 
\end{abstract}

\body

\section{Introduction}
\label{Intro}

\subsection{Superfluidity in nuclei and BCS ansatz}
\label{pairing_intro}

The structure of the nucleus and properties of extended nuclear systems strongly depend on their possible superfluid
nature~\cite{dean03a}. In nuclei, pairing impacts all low-energy properties of the system, e.g. masses, separation energies, radii as well as individual, rotational and vibrational excitation modes. The role of pairing correlations is emphasized close to the drip-lines due to the proximity of the Fermi surface to one- and two-particle emission thresholds~\cite{doba03}. In neutron stars, superfluidity also plays a key role, e.g. it impacts post-glitch timing observations~\cite{Avogadro:2006ed} or their cooling history~\cite{heiselberg3}.

Despite the major role played by pairing in nuclei, its microscopic understanding is rather poor. Given realistic inter-nucleon forces, questions of interest include (i) how much pairing is due to the direct attraction between nucleons on the one hand and how much it is due to the indirect coupling resulting from the exchange of medium fluctuations on the other, (ii) what impact Coulomb and three-nucleon forces have.

The key feature of any method based on the Bardeen-Cooper-Schriffer (BCS) ansatz is to address explicitly the non-perturbative process associated with the formation of quasi-bound (Cooper) pairs within the many-fermion system. This however done at the price of breaking the $U(1)$ symmetry associated with particle-number conservation, which eventually calls for its restoration. The latter is of particular relevance to mesoscopic systems such as the atomic nucleus. In an ab-initio context, e.g., within the frame of many-body perturbation theory~\cite{mehta1,henley} or self-consistent Green's function theory~\cite{gorkov}, the BCS rationale is incorporated by expanding the exact solution around an unperturbed state of the BCS or Bogoliubov type that already captures the key non-perturbative physics responsible for superfluidity.

\subsection{Low-momentum inter-nucleon interactions}
\label{lowk}

Establishing a realistic nuclear Hamiltonian, which is the basic precursor to any ab-initio many-body calculations, is a challenge for low-energy nuclear physics.  The two-body sector has been intensively investigated and various interactions exist that reproduce nucleon-nucleon scattering phase shifts with $\chi^2/N_{\mathrm{dof}} \approx 1$ in the elastic regime (up to about $300\mbox{--}350$\,MeV energy in the laboratory frame).  The unsettled frontier is three- and higher-body forces~\cite{Machleidt:2010kb,KalantarNayestanaki:2011wz,epelbaum07b}.

The development of chiral effective field theory ($\chi$-EFT) has made possible~\cite{Epelbaum:2008ga} to connect low-energy inter-nucleon forces to underlying Quantum Chromo Dynamics (QCD), whose relevant high-energy effects are renormalized through fitted low-energy contact terms~\cite{walzl01,epelbaum06,Epelbaum:2008ga}. The main benefits of $\chi$-EFT are (i) to formulate the problem at hand in terms of relevant low-energy degrees of freedom (pions and nucleons) while retaining the (chiral) symmetry (breaking) of the underlying theory (QCD), (ii) to provide a consistent building of all relevant operators and (iii) to naturally explain the phenomenologically-observed hierarchy that makes two-nucleon (2N) interactions more important than three-nucleon (3N) interactions, which themselves dominate four-nucleon (4N) forces etc. Such a hierarchy relates to a power counting that organizes the infinite set of diagrams in the $\chi$-EFT Lagrangian~\cite{Epelbaum:2008ga} according to their scaling with $(Q/\Lambda_{\chi})^{\nu}$, where $Q$ embodies typical low-momentum processes and degrees of freedom at play while $\Lambda_{\chi}$, the so-called chiral-symmetry-breaking scale, denotes the hard scale characterizing omitted degrees of freedom and driving low-energy constants in the Lagrangian. As such, $\chi$-EFT underlines that any nuclear Hamiltonian is effective and take the $\Lambda_{\chi}$-dependent form
\begin{equation}
H = T + V^{\text{2N}} + V^{\text{3N}} + \ldots
\end{equation}
where $V^{\text{2N}}$ first contributes at leading order (LO) while $V^{\text{3N}}$ only enters at next-to-next-to leading order (N$^{2}$LO). 

Such nuclear Hamiltonians display several sources of non-perturbative behaviour that complicate many-body calculations. First are the strong virtual coupling between low- and high-momentum\footnote{In the present context of $\chi$-EFT, "high-momenta" remain below $\Lambda_{\chi}$.} modes driven by the non-observable short-range part of central and tensor forces. The second source of non-perturbative behaviour relates to the existence of weakly- and nearly-bound two-body states associated with $^3S_1$ and $^1S_0$ partial-waves of the 2N interaction, respectively. Nearly-bound neutron-neutron and proton-proton states in the vacuum are the precursor of Cooper pairs that emerge at finite nuclear densities and that are central to the present discussion.

Progress toward controlled calculations has long been hindered by the difficulty to solve the nuclear many-body problem in presence of the strong virtual coupling between low- and high-momentum modes. This has historically been accepted as an unavoidable reality. Recently, EFT and renormalization group (RG) methods~\cite{Bogner:2003wn,Bogner:2009bt} have promoted a different view point based on the fact that the Hamiltonian (potential) is not an observable to be fixed from experiment. There rather exists an infinite number of Hamiltonians (potentials) capable of accurately describing the low-energy physics~\cite{Lepage:1997cs}. Starting from, e.g., a $\chi$-EFT Hamiltonian, one can take advantage of such a freedom to perform a (unitary) transformation $U(\Lambda)$ with the aim of {\it decoupling} low-momentum modes from high-momentum ones. Doing so corresponds to keeping the physics (i.e. any true observable) invariant while transforming the Hamiltonian according to
\begin{eqnarray}
H(\Lambda) &\equiv& U(\Lambda)H U^{\dagger}(\Lambda)\equiv T + V^{\text{2N}}(\Lambda) + V^{\text{3N}}(\Lambda) + \ldots \, .
\end{eqnarray}
The RG transformation "lowers the resolution scale $\Lambda$" ($<\Lambda_{\chi}$) of the Hamiltonian while preserving the original truncation error. Just as they depended on the original resolution scale $\Lambda_{\chi}$, 2N, 3N,\ldots AN interactions further depend on $\Lambda$ while  observables do not~\cite{Bedaque:2002mn,Bogner:2009bt,Epelbaum:2008ga}.  The main benefit of the above transformation is that each component of $H(\Lambda)$ typically becomes {\it softer} as $\Lambda$ is lowered such that many-body calculations become more perturbative as far as the virtual coupling to high-momenta is concerned~\cite{Bogner:2005sn}. One must however note that the source of non-perturbative physics associated with weakly- and nearly-bound two-body states, which is the focus of the present contribution, remains untouched by the lowering of $\Lambda/\lambda$. Even though Cooper pairs are tamed down as the density of the medium increases~\cite{Ramanan:2007bb}, they  must be explicitly dealt with through non-perturbative techniques at and below nuclear saturation density. As discussed above, it is a key virtue of BCS-based methods to already account for the dominant effects of such a non-perturbative process through a simple zero-order ansatz.

Other practical advantages of proceeding to a RG transformation over a reasonable range of $\Lambda$ values relate to the fact that the unitarity of the transformation is not significantly compromised (and thus the physics not significantly altered) (i) by omitting induced many-body forces beyond a certain rank and (ii) by eventually truncating the size of the initial  Hilbert space. Of course, the improved convergence pattern deriving from the latter truncation and the preserved hierarchy of AN forces built into $H$ must be thoroughly checked in actual many-body calculations~\cite{Jurgenson:2009qs,Anderson:2010aq,Roth:2011ar,Jurgenson:2010wy}. As a matter of fact, one can exploit the change of many-body observables with $\Lambda$ to study the underlying physics scales and evaluate the incompleteness of approximate calculations or the impact of dropping many-body forces in the transformed Hamiltonian $H(\Lambda)$.

%\begin{figure}[ht]
%\centerline{
%  \subfigure[]
%     {\epsfig{figure=figures/srg_schematic.eps,width=1.8in}\label{ra_fig2b}}
%  \hspace*{4pt}
%  \subfigure[]
%     {\epsfig{figure=figures/vlowk_schematic.eps,width=1.8in}\label{ra_fig2a}}
%}
%\caption{Schematic illustration of two types of RG evolution for 2N potentials displayed as matrices in (relative) momentum space: (a)~SRG running in $\lambda$, and (b)~$\vlowk$ running in $\Lambda$. At each $\Lambda$ or $\lambda$, the matrix elements outside of the corresponding lines are zero, so that high- and low-momentum states are being decoupled. Taken from Ref.~\cite{Bogner:2009bt}.}\label{fig:schematic}
%\end{figure}

Two main classes of RG transformations are used to construct low-momentum interactions. For a schematic illustration of the evolution of 2N potentials' matrix elements in (relative) momentum space within each of this two methods, see Fig. 9 of Ref.~\cite{Bogner:2009bt}. In the first approach, denoted as similarity renormalization group (SRG), the decoupling is achieved through a unitary transformation over the Hilbert (Fock) space defined in connection with the original, e.g., $\chi$-EFT Hamiltonian. A momentum scale $\lambda$ that measures the extent of off-diagonal coupling is lowered through flow equations such that the potentials are driven toward a band diagonal form in momentum space. In the second approach, generically referred to as the $\vlowk$ approach, the decoupling is achieved by a transformation that is in fact not unitary over the original Hilbert (Fock) space. The renormalized potential is such that matrix elements beyond the lowered momentum cutoff $\Lambda$ are set to zero. In practice, the low-energy part of the renormalized potentials are eventually very similar in both approaches such that the breaking of unitary in the $\vlowk$ approach has no influence on low-energy observables~\cite{Bogner:2009bt}.

\subsection{Outline}
\label{outline}

The present contribution is organized as follows. Section~\ref{OEMS} discusses the extraction of the most basic observable related to pairing in nuclei. Section~\ref{EDF} reviews the recent use of realistic low-momentum interactions within the frame of the semi-empirical single-reference energy density functional (SR-EDF) method. Section~\ref{Gorkov} reports on the even more recent implementation of ab-initio self-consistent Gorkov Green's function calculations based on low-momentum interactions that constitutes a path towards a fully microscopic description of superfluidity in nuclei.

\section{Pairing information from the odd-even mass staggering}
\label{OEMS}

The present contribution focuses on the most basic information related to pairing, i.e. the "pairing gap" extracted via three-point mass differences
\begin{equation}
\label{eq:3pmass}
\Delta^{(3)}(N) \equiv \frac{(-1)^N}{2} \left[E_0^{N+1} - 2E_0^{N} + E_0^{N-1}\right] \: ,
\end{equation}
where $N$ and $E_0^{N}$ denote the number of nucleons and the ground-state binding energy, respectively. Extracting $\Delta^{(3)}(N)$ is motivated by the relation between the odd-even staggering of nuclear masses and the lack of binding of odd-even systems due to the presence of an unpaired nucleon~\cite{bohr69a}. As such, $\Delta^{(3)}(N)$ is indeed dominated by the "pairing gap". Still, finite-difference formulas are contaminated by sub-leading contributions that are not related to the pairing gap~\cite{duguet02a,duguet02b} (see Eq.~\ref{eq:3pmass-th} below). Consequently, and to avoid any mismatch when comparing theory to experiment, it is recommended, whenever possible, to perform differences of actual theoretical masses rather than to use an approximate formula for the theoretical gap.

\section{Semi-empirical energy density functional calculations}
\label{EDF}

\subsection{Elements of formalism}

The SR-EDF method implements the breaking of $U(1)$ symmetry and takes the form of an {\it effective} Hartree-Fock-Bogoliubov (HFB)
formalism~\cite{bender03b}. The total energy is postulated under the form a functional $E^{N}_0=\mathcal{E}[\rho,\kappa,\kappa^{\ast}]$ of the (symmetry breaking)
one-body density $\rho$ and pairing tensor $\kappa$ computed from an {\it
auxiliary} product state $|\Phi_0\rangle$ of the Bogoliubov type,
\begin{equation}
\rho_{ab} \equiv \langle\Phi_0| c^\dagger_{b} \,c_a | \Phi_0\rangle
 \, , ~~~~~~~~~~~~~
\kappa_{ab} \equiv \langle\Phi_0 | c_b \, c_a | \Phi_0\rangle \, ,
\end{equation}
where $\{c^\dagger_a\}$ denotes an arbitrary basis of the one-body Hilbert space $\mathcal{H}_1$. The minimization of
$\mathcal{E}[\rho,\kappa,\kappa^{\ast}]$ under the constrain that $N = \langle  \Phi_0 |  N |  \Phi_0 \rangle$, leads to solving effective Bogoliubov-De-Gennes equations~\cite{ring80a}
\begin{equation}
\left(
  \begin{array}{cc}
  h-\lambda   & \Delta  \\
-\Delta^{\ast}    &-h^{\ast}+\lambda
  \end{array} \right) \,
  \left(
  \begin{array}{c}
 \mathcal{U}  \\
\mathcal{V}
  \end{array} \right)_{\mu}
           = E_{\mu}  \,   \left(
  \begin{array}{c}
 \mathcal{U}  \\
\mathcal{V}
  \end{array} \right)_{\mu}
 \, \, \, , \label{HFBeigenvaluekappa}
\end{equation}
where $(\mathcal{U}, \mathcal{V})_{\mu}$ are the upper and lower components of Bogoliubov quasi-particle
eigenstates whereas $E_{\mu}$ denotes the corresponding quasi-particle energies. The
single-particle $(h)$ and pairing $(\Delta)$ fields are defined through
\begin{subequations}
\label{fields}
\begin{eqnarray}
h_{ab} &\equiv& \frac{\delta \mathcal{E}}{\delta \rho_{ba}}
  \equiv   t_{ab}+\Sigma^{11}_{ab}(\text{eff})  \equiv t_{ab}+\sum_{cd} \overline{v}^{ph}_{acbd} \; \rho_{dc} , \\ 
\Delta_{ab} &\equiv& \frac{\delta \mathcal{E}}{\delta \kappa^{\ast}_{ab}}
    \equiv \Sigma^{12}_{ab}(\text{eff}) \equiv \frac{1}{2}
        \sum_{cd} \overline{v}^{pp}_{abcd}\; \kappa_{cd} \, \, \, , \label{fieldsdef}
\end{eqnarray}
\end{subequations}
where $t_{ab}$ denotes the matrix elements of the kinetic energy operator. In Eq.~\ref{fields}, {\it effective} normal $\Sigma^{11}(\text{eff})$ and anomalous $\Sigma^{12}(\text{eff})$ self-energies, as well as effective particle-hole $\overline{v}^{ph}$ and particle-particle $\overline{v}^{pp}$ kernels, are introduced for interpretation purposes. The effective character of the EDF approach relates to the fact that $\mathcal{E}[\rho,\kappa,\kappa^{\ast}]$ and $\Sigma^{gg'}(\text{eff})$ are meant to re-sum correlations that go largely beyond Hartree, Fock and Bogoliubov diagrams calculated in terms of vacuum inter-nucleon interactions.

Modern empirical parametrizations of existing, e.g. Skyrme or Gogny, EDFs provide a fair description of bulk and spectroscopic properties of known nuclei~\cite{bender03a}. However, they lack predictive power away from available data and a true spectroscopic quality, especially regarding the part that drives superfluidity. Consequently, efforts are currently made to empirically improve the analytical form and the fitting procedure of functionals, e.g. see Refs.~\cite{Margueron:2007uf,Yamagami:2008ks,Chamel:2010rw} for recent attempts to pin down the isovector content of local pairing functionals. 

Along with improving the phenomenology at play, it is relevant to understand the processes responsible for superfluidity in nuclei in microscopic terms, i.e. starting from vacuum inter-nucleon interactions. The lowest-order contribution to the two-particle irreducible pairing kernel $\overline{v}^{pp}$ is provided by vacuum interactions themselves, while higher-order terms include the induced interaction associated with the exchange of collective medium fluctuations between paired particles~\cite{barranco04a,gori05,Pastore:2008zi,Idini:2011zz}. A fundamental, yet unresolved, question relates to how much of the pairing gap in finite nuclei is accounted for at lowest order and how much is due to higher-order processes. 

Awaiting for fully ab-initio calculations of mid-mass nuclei (see Sec.~\ref{Gorkov}), we have recently set up a semi-empirical SR-EDF scheme that combines empirical parametrizations of the single-particle field $h$ with a pairing field $\Delta$ built at first order in 2N~\cite{duguet04a,Duguet:2007be,Lesinski:2008cd,Hebeler:2009dy,Baroni:2009eh} (including Coulomb~\cite{Lesinski:2008cd}) and 3N~\cite{Duguet:2010qw,Lesinski:2011rn} low-momentum interactions. The objective was to provide a partial answer to the fundamental question alluded to above. The next section summarizes the main results obtained following such an approach.

\subsection{Calculation scheme}
\label{implementationEDF}

Our calculations start from the N$^3$LO 2N potential (EM $500$\,MeV) of Ref.~\cite{entem03}, which is then RG-evolved using the $V_{{\rm low}\,k}$ procedure and a smooth $n_{\rm exp}=4$ regulator~\cite{bogner07a} down to $\Lambda = 1.8 - 2.8$\,fm$^{-1}$. We complement it~\cite{Hebeler:2010xb} at each $\Lambda$ by the leading N$^2$LO 3N force, which is then averaged over the third nucleon following the procedure detailed in Ref.~\cite{Lesinski:2011rn}. Adding both contributions, we obtain $\overline{v}^{pp}$ at first order in vacuum interactions for various $\Lambda$ values. To be consistent, the isoscalar and isovector effective masses of the Skyrme parametrization used to build $\overline{v}^{ph}$ is constrained from Hartree-Fock calculations of neutron and symmetric nuclear matter based on the same low-momentum 2N and 3N interactions~\cite{Hebeler:2009dy}. Last but not least, odd-even nuclei are computed through the self-consistent blocking procedure performed within the filling approximation~\cite{perezmartin08a,Duguet:2010qw} such that $\Delta^{(3)}(N)$ is indeed extracted through total mass differences. 

\subsection{Results}
\label{resultsEDF}

\begin{figure*}[t]
%\begin{center}
%\includegraphics[scale=0.47,clip=]{figures/gaps-d3th-nnn-or-not-new.eps}
\hspace{-1.5cm}
\includegraphics[scale=0.72,clip=]{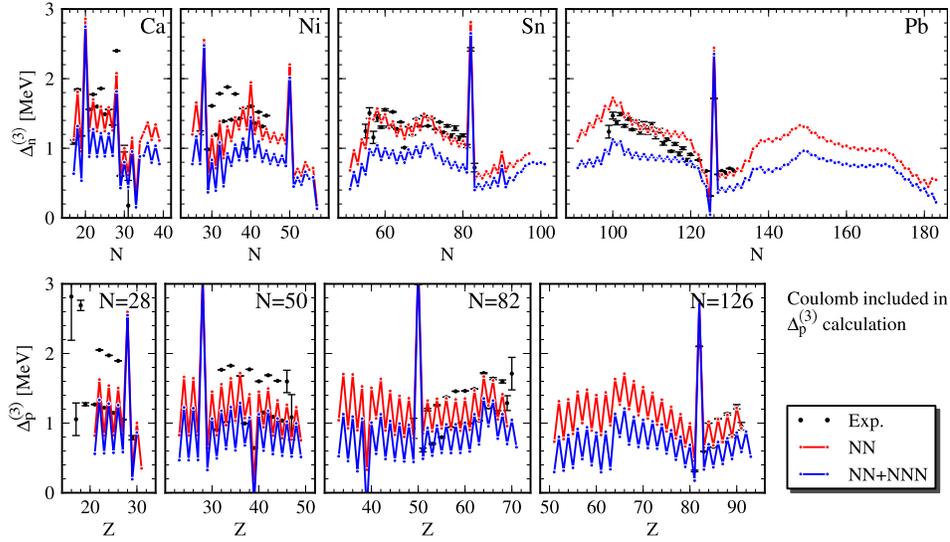}
%\end{center}
\vspace*{-5mm}
\caption{Theoretical and experimental $\Delta^{(3)}(N)$ along isotopic/isotonic chains based on
low-momentum 2N and 3N interactions with $\Lambda/\Lambda_{\rm 3NF}
=1.8/2.0$\,fm$^{-1}$. \label{fig:gaps}}
\end{figure*}

Figure~\ref{fig:gaps} compares theoretical and experimental three-point mass differences along several semi-magic isotopic/isotonic chains. Results obtained with and without 3N contributions to $\overline{v}^{pp}$ are displayed. Using the 2N contribution only, neutron and proton pairing gaps are close to experimental ones for a large set of semi-magic spherical nuclei, although experiment is underestimated in the lightest systems. The addition of the first-order 3N contribution lowers pairing gaps systematically by about $30\%$. 

In Ref.~\cite{Lesinski:2011rn}, the $\Lambda$ dependence of the pairing gaps was also studied as a way to estimate
uncertainties due to short-range higher-order AN forces and to the incomplete many-body treatment. The analysis led to a theoretical uncertainty of order $\approx 100-250 \, {\rm keV}$ in all semi-magic chains investigated.

All in all, the recent studies reported in Refs.~\cite{Duguet:2007be,Lesinski:2008cd,Hebeler:2009dy,Baroni:2009eh,Duguet:2010qw,Lesinski:2011rn} have supported the following conclusions: (i) Coulomb repulsion is responsible for a significant proton anti-pairing effect and must be explicitly incorporated in the pairing kernel, (ii) it is essential to include the 3N contributions to the pairing kernel to reach a quantitative description of nuclear pairing gaps, (iii) first-order results leave about $30 \%$ room for contributions from higher orders, e.g. from the coupling of Cooper pairs to (collective) density, spin and isospin fluctuations, which is consistent with phenomenological calculations showing that induced interactions are overall attractive in nuclei ~\cite{barranco04a,gori05,Pastore:2008zi,Idini:2011zz},
(iv) future investigations need to also compute the normal self-energy and higher-order contributions to the pairing kernel consistently from low-momentum 2N and 3N interactions. This is the objective of the next section to report on the first step in that direction.

\section{Ab-initio self-consistent Gorkov-Green's function theory}
\label{Gorkov}

The ultimate goal is to perform fully microscopic calculations of pairing properties in mid- and heavy-mass nuclei. One way to do so consists of performing ab-initio self-consistent Gorkov Green's function (SCGGF) calculations~\cite{gorkov} based on low momentum interactions. While it is exact in the limit where self-energy diagrams are summed to all orders, SCGGF reduces to the (ab-initio) HFB approximation at lowest order. In practice, calculations are the result of a compromise, i.e. one must implement a tractable truncation scheme that approximates the exact solution well enough. In that respect, switching from conventional hard-core potentials to low-momentum interactions is instrumental as it makes finite-order schemes qualitatively and quantitatively viable~\cite{Hebeler:2010xb}. 

Recently, V. Som\`a, C. Barbieri and myself have implemented SCGGF theory at second order on the basis of low-momentum 2N interactions~\cite{soma11a,soma11b}. These are the first ever ab-initio calculations of their kind in finite nuclei. They constitute a first step towards a quantitative investigation of induced interaction effects from a purely ab-initio perspective. To be of quantitative interest, such calculations need to be extended to 3N interactions and to more advanced truncation schemes allowing for the coupling of the Cooper pair to {\it collective} fluctuations. These constitute our mid-term objectives.

\subsection{Elements of formalism}

Let basis $\{c_a^{\dagger} \}$ splits into two blocks that can be mapped onto each other through time-reversal. Let us then introduce a partner basis $\{\bar{c}_a^{\dagger} \}$ through
\begin{equation}
\label{eq:gen_aad}
\bar{c}_{a}^{\dagger}(t) \equiv \eta_a c_{\bar{a}}^{\dagger}(t)\, , \qquad
\bar{c}_{a}(t) \equiv \eta_a c_{\bar{a}}(t) \: ,
\end{equation}
which corresponds to exchanging the state $a$ by its time-reversal partner $\bar{a}$ up to the phase $\eta_a$. By convention $\bar{\bar{a}}=a$ and $\eta_a \, \eta_{\bar{a}}~=~-1$.

In Gorkov formalism, one targets the ground state $|  \Psi_0 \rangle$ of the grand-canonical-like potential $\Omega = H - \mu N$, where $\mu$ is the chemical potential and $N$ the particle-number operator, having the number $N = \langle  \Psi_0 |  N |  \Psi_0 \rangle$ of particles on average. In order to access the complete one-body information contained in $| \Psi_0 \rangle$, one must introduce a set of four Green's functions, known as Gorkov propagators~\cite{gorkov}. Defining an "annihilation" column vector and a "creation" row vector through
\begin{equation}
\label{eq:A_nambu}
\nC_a(t) \equiv
\left(
\hspace{-.2cm}
\begin{tabular}{c}
$c_a(t)$  \\
$\bar{c}_a^{\dagger}(t)$
\end{tabular}
\hspace{-.2cm}
\right) \: ,
 \qquad
\nC_a^{\dagger}(t) \equiv
\left(
\hspace{-.2cm}
\begin{tabular}{cc}
$c_a^{\dagger}(t)$  &
$\bar{c}_a(t)$
\end{tabular}
\hspace{-.2cm}
\right) \: ,
\end{equation}
one can write the four propagators in matrix representation~\cite{Nambu:1960tm} through
\begin{eqnarray}
\label{eq:gnambu}
i \, \nG_{ab}(t,t') &\equiv&
\langle \Psi_0 | T \left \{
\nC_{a}(t) \nC_{b}^{\dagger}(t')
\right\}
| \Psi_0 \rangle
=
i \, \left(
\begin{tabular}{cc}
$G^{11}_{ab}(t,t')$ & $G^{12}_{ab}(t,t')$ \\
& \\
$G^{21}_{ab}(t,t')$ & $G^{22}_{ab}(t,t')$
\end{tabular}
\right) \: .
\end{eqnarray}
Self-consistent, i.e. {\it dressed}, propagators are solution of Gorkov's equation
\begin{equation}
\label{eq:eigen_uv_bar}
\left.
\left(
\begin{tabular}{cc}
$T  + \Sigma^{11}(\omega)- \lambda$ & $\Sigma^{12}(\omega)$ \\
$\Sigma^{21}(\omega)$ & $-T + \Sigma^{22}(\omega) + \lambda$
\end{tabular}
\right)
\right|_{\omega_k}
\left(
  \begin{array}{c}
 \mathcal{U}  \\
\mathcal{V}
  \end{array} \right)_{\mu}=\omega_{\mu} \,
\left(
  \begin{array}{c}
 \mathcal{U}  \\
\mathcal{V}
  \end{array} \right)_{\mu} \: ,
\end{equation}
whose outputs are Gorkov's amplitudes  $(\mathcal{U}, \mathcal{V})_{\mu}$ and corresponding pole energies $\omega_{\mu}$, in terms of which the four propagators can be expressed~\cite{soma11a}. Equation~\ref{eq:eigen_uv_bar} generalizes Eq.~\ref{HFBeigenvaluekappa} in the sense that normal ($\Sigma^{11}(\omega)$ and $\Sigma^{22}(\omega)$) and anomalous ($\Sigma^{12}(\omega)$ and $\Sigma^{21}(\omega)$) irreducible self-energies act here as {\it energy-dependent} potentials. At first order in vacuum interactions, Eq.~\ref{eq:eigen_uv_bar} reduces to an ab-initio\footnote{This is at variance with the {\it effective} character of Eq.~\ref{HFBeigenvaluekappa}  in which {\it energy-independent} fields $h$ and $\Delta$ are meant to effectively account for correlations that go beyond strict Hartree, Fock and Bogoliubov diagrams.} HFB equation with normal and anomalous self-energies accounting for Hartree-Fock and Bogoliubov diagrams, respectively. Proceeding to an actual calculation relates to truncating the diagrammatic expansion of the self-energies $\Sigma^{gg'} (\omega)$. As opposed to perturbation theory, the expansion involves skeleton diagrams expressed in terms of dressed propagators solution of Eq.~\eqref{eq:eigen_uv_bar}. This key feature of {\it self-consistent} Green's function methods allows the re-summation of self-energy insertions to all orders and makes the method intrinsically non-perturbative and iterative. Eventually, the total energy is computed via the Koltun-Galitskii sum rule~\cite{Koltun:1972}
\begin{eqnarray}
\label{eq:koltun_gorkov}
E^{N}_0 &=&
\sum_{ab} \frac{i}{4 \pi} \int_{C \uparrow} d \omega \, G^{11}_{ba} (\omega) \left[ t_{ab}
+ \omega \, \delta_{ab} \right] \, ,
\end{eqnarray}
where the integration is performed over a closed contour in the upper half of the complex plane.  Extensive details regarding both the formalism and the computational scheme can be found in Refs.~\cite{soma11a,soma11b}.

Computing odd-even nuclei requires in the present context to perform Gorkov calculations for a state $| \Psi_0 \rangle$ having an odd number-parity quantum number. This is however beyond the scope of the present work. The next best approximation consists in keeping an even number-parity state while accounting for the self-consistent blocking of a quasi-particle within the filling approximation~\cite{perezmartin08a}. Such an approximation remains however to be formulated within the frame of Gorkov-Green's function formalism. Consequently, the next best estimate of the ground-state energy of an odd-even system is obtained through~\cite{duguet02a,duguet02b}
\begin{equation}
\label{eq:oddE}
E_0^{N} \approx E_0^{N \, *} + \omega^{N}_{F} \: ,
\end{equation}
where $E_0^{N \, *}$ is the energy of the odd nucleus computed as it were an even one, i.e. as a fully paired even number-parity state having an odd number of particles on average, while $\omega^{N}_{F}$ denotes the lowest pole energy extracted from Eq.~\ref{eq:eigen_uv_bar} for that calculation. For even $N$ one simply has $E_0^{N} = E_0^{N \, *}$, i.e. $E_0^{N \, *}$ provides the energy curve on which both even and odd nuclei would lie in the absence of odd-even mass staggering. With such a decomposition of the energy, Eq.~\ref{eq:3pmass} becomes
\begin{equation}
\label{eq:3pmass-th}
\Delta^{(3)}(N) \approx \frac{(-1)^N}{2} \frac{\partial^2 E_0^{N \, *}}{\partial N^2} + \Delta_{F}(N) \: .
\end{equation}
The second derivative of $E_0^{N \, *}$ is smooth with $N$ but provides $\Delta^{(3)}(N)$ with a rapidly oscillating contribution\footnote{See Figs.~\ref{fig:gaps}.} because of the factor $(-1)^N$ that comes with it in Eq.~\eqref{eq:3pmass-th}. Clearly, such a contribution is not related to the pairing gap. The second contribution to $\Delta^{(3)}(N)$ relates specifically to the unpaired character of the odd nucleon and does extract, in open-shell nuclei, the pairing gap at the Fermi energy~\cite{duguet02a,duguet02b}
\begin{eqnarray}
\label{gapfermi}
\Delta_{F}(N)  \equiv
\left\{
\begin{tabular}{cl}
$\omega^{N}_{F}$ & \,\, for $N$ odd \\
$(\omega^{N-1}_{F}+\omega^{N+1}_{F})/2$ & \,\,  for $N$ even
\end{tabular}
\right.
\: .
\end{eqnarray}

\subsection{Calculation scheme}

We performed self-consistent second-order calculations~\cite{soma11b}, i.e. first- and second-order diagrams are included in the computation of the self-energies (see Figs.~1-3 of Ref.~\cite{soma11a}), of Ca isotopes. Such a truncation scheme constitutes a Kadanoff-Baym $\Phi$-derivable approximation, which automatically ensures the exact fulfilment of conservation laws~\cite{baym62}. The $\chi$-EFT 2N potential~\cite{entem03} is RG-evolved down to $\Lambda=2.1$\,fm$^{-1}$ using the $V_{{\rm low}\,k}$ procedure and a sharp regulator. In addition to the direct (i.e. first order) contribution to the pairing kernel incorporated (along with the corresponding diagram from the 3N interaction) in the semi-empirical EDF calculations reported on in Sec.~\ref{EDF}, the present calculation includes the coupling of the Cooper pair to {\it non-collective} density, spin and isospin fluctuations. Such a test case calculation is performed on a single-processor and uses a restricted model space of 7 major harmonic oscillator shells. As such, results presented here must only be taken as indicative. Converged calculations require a multi-processors architecture and will be reported on in Ref.~\cite{soma11b}.

%\begin{figure}[h]
%\begin{center}
%\begin{tabular}{ccc}
%\nlfirstn & \nlsecondna & \nlsecondnb \\
%\end{tabular}
%\end{center}
%\caption{First- and second-order contributions to the normal self-energy $\Sigma^{11}(\omega)$. Double lines denote self-consistent normal (two arrows in the same direction)  and anomalous (two arrows in opposite directions) propagators.}
%\label{fig:first}
%\end{figure}

%\begin{figure}[h]
%\begin{center}
%\begin{tabular}{ccc}
%\nlfirstaa & \nlsecondaaanew &  \nlsecondabbnew \\
%\end{tabular}
%\end{center}
%\caption{Same as Fig.~\ref{fig:first} for the anomalous self-energy $\Sigma^{21}(\omega)$.}
%\label{fig:seconda}
%\end{figure}

\subsection{Results}

\begin{figure*}[t]
\begin{center}
\includegraphics[scale=0.38,clip=]{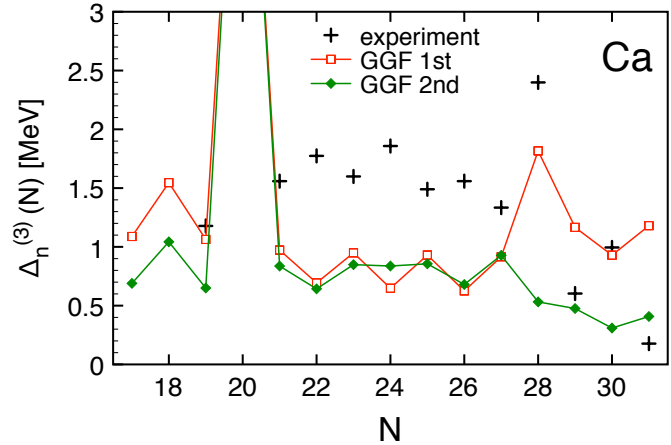}
\end{center}
\vspace*{-5mm}
\caption{Neutron three-point mass difference in Ca isotopes. Theoretical values are from first and second-order SCGGF calculations based on
a low-momentum 2N interaction with $\Lambda =2.1$\,fm$^{-1}$. See text for details.\label{gapCaGorkov}}
\end{figure*}

Figure~\ref{gapCaGorkov} compares theoretical and experimental three-point mass differences from $^{37}$Ca to $^{51}$Ca. Results obtained at first and second-order are displayed. We observe three main features (1) the second-order contribution to normal and anomalous self-energies generates a slight decrease of $\Delta^{(3)}(N)$, (2) gaps at first and second order account for about half of experimental values, (3) the staggering of $\Delta^{(3)}(N)$ is {\it inverted} compared to experiment. All such features will be discussed in details in Ref.~\cite{soma11b} on the basis of better converged calculations performed using larger single-particle model spaces. Let us however make a few tentative comments regarding each of these three points.
\begin{enumerate}
\item The coupling to non-collective fluctuations seems to have little influence on $\Delta^{(3)}(N)$ and to have the tendency to slightly suppress it. Although one should wait for converged calculations and for a thorough analysis of the result before drawing any conclusion, the fact that the collective character of the fluctuations the Cooper pair couples plays a key role is expected~\cite{barranco04a,gori05,Pastore:2008zi,Idini:2011zz}. Treating collective fluctuations within the present ab-initio setting is a challenge that will be addressed in the mid-term future. 
\item The omission of the 3N force in the computation of the normal self-energy results in a significantly too small effective mass. This prevents the pairing kernel from fully expressing its strength and eventually generates too low gaps. The second-order contribution to the normal self-energy from the 2N interaction does not compensate (enough) for this defect such that incorporating the 3N force contribution is mandatory to obtain a realistic effective mass and meaningful predictions of the pairing gaps\footnote{In the context of EDF calculations discussed in Sec.~\ref{EDF},  the realistic effective mass is obtained empirically by fitting appropriate free parameters of the Skyrme EDF.}. It particular, having a realistic effective mass will feedback on the second-order contribution to the anomalous self-energy discussed in point (1) in a way that remains to be seen. Note that the indirect enhancement of the gaps associated with the 3N force contribution to the normal self-energy will counteract its repulsive contribution to the pairing kernel discussed in Sec.~\ref{resultsEDF}.
\item The inverted staggering of $\Delta^{(3)}(N)$ emphasizes (see first term of Eq.~\ref{eq:3pmass-th}) the wrong curvature of $E_0^{N \, *}$ obtained in the present calculation as one adds neutrons~\cite{soma11d}. Such a feature relates to the lack of saturation of nuclear matter, and correspondingly to a wrong asymmetry energy, generated by low-scale 2N interactions when omitting the corresponding 3N interaction~\cite{Hebeler:2010xb}. Within each shell, the energy per particle increases as one adds neutrons whereas it should decrease. Increasing the scale $\Lambda$ of the 2N interaction partly corrects for such a wrong pattern. Still, resolving this issue satisfactorily necessitates, independently of the value of $\Lambda$, the explicit treatment of 3N interactions\footnote{The present issue relates again to the contribution of the 3N force to the {\it normal} self-energy, not the anomalous one.} within the present ab-initio setting. This again constitutes one of our two mid-term objectives.
\end{enumerate}

\section{Conclusions}
\label{conclusions}

The present contribution has reviewed recent advances made toward a microscopic understanding of superfluidity in nuclei using many-body methods based on the BCS ansatz and low-momentum inter-nucleon interactions. Steps towards a truly ab-initio treatment of pairing in medium-mass nuclei  have been briefly highlighted.

\section{Acknowledgments}

I wish to thank C. Barbieri, K. Bennaceur, K. Hebeler, T. Lesinski, J. Meyer, A. Schwenk and V. Som\`a very warmly for the fruitful collaborations that have generated the scientific advances summarized in the present contribution. I particularly wish to thank V. Som\`a and C. Barbieri for letting me use some of our yet to be published results. I finally thank T. Lesinski and V. Som\`a for having prepared the two figures used in the present contribution.

\bibliographystyle{ws-rv-van}
\bibliography{duguet_pairing}

\begin{thebibliography}{53}
\providecommand{\natexlab}[1]{#1}
\providecommand{\url}[1]{\texttt{#1}}
\expandafter\ifx\csname urlstyle\endcsname\relax
  \providecommand{\doi}[1]{doi: #1}\else
  \providecommand{\doi}{doi: \begingroup \urlstyle{rm}\Url}\fi

\bibitem{dean03a}
D.~J. Dean and M.~Hjorth-Jensen, \emph{Rev. Mod. Phys.} {\bf 75}, \penalty0
  607,  (2003).

\bibitem{doba03}
J.~Dobaczewski and W.~Nazarewicz, \emph{Prog. Theor. Phys. Suppl.} {\bf 146},
  \penalty0 70,  (2003).

\bibitem{Avogadro:2006ed}
P.~Avogadro, F.~Barranco, R.~A. Broglia, and E.~Vigezzi, \emph{Phys. Rev.} {\bf
  C75}, \penalty0 012805,  (2007).

\bibitem{heiselberg3}
H.~Heiselberg and M.~Hjorth-Jensen, \emph{Phys. Rep.} {\bf 328}, \penalty0 237,
   (2000).

\bibitem{mehta1}
M.~L. Mehta.
\newblock PhD thesis,  (1961).
\newblock CEA/Saclay/DSM/SPhT Report No. T61/0134.

\bibitem{henley}
E.~M. Henley and L.~Wilets, \emph{Phys. Rev.} {\bf 133}, \penalty0 B1118,
  (1964).

\bibitem{gorkov}
L.~P. Gorkov, \emph{Sov. Phys. JETP}. {\bf 34}, \penalty0 505,  (1958).

\bibitem{Machleidt:2010kb}
R.~Machleidt and D.~Entem, \emph{J.Phys.G}. {\bf G37},  (2010).

\bibitem{KalantarNayestanaki:2011wz}
N.~Kalantar-Nayestanaki, E.~Epelbaum, J.~Messchendorp, and A.~Nogga,
  \emph{Rept.Prog.Phys.} {\bf 75}, \penalty0 016301,  (2012).

\bibitem{epelbaum07b}
E.~Epelbaum, \emph{Eur. Phys. J. A}. {\bf 34}, \penalty0 197,  (2007).

\bibitem{Epelbaum:2008ga}
E.~Epelbaum, H.-W. Hammer, and U.-G. Meissner, \emph{Rev. Mod. Phys.} {\bf 81},
  \penalty0 1773,  (2009).

\bibitem{walzl01}
M.~Walzl, U.-G. Mei{\ss}ner, and E.~Epelbaum, \emph{Nucl. Phys.} {\bf A693},
  \penalty0 663,  (2001).

\bibitem{epelbaum06}
E.~Epelbaum, \emph{Prog. Part. Nucl. Phys.} {\bf 57}, \penalty0 654,  (2006).

\bibitem{Bogner:2003wn}
S.~K. Bogner, T.~T.~S. Kuo, and A.~Schwenk, \emph{Phys. Rept.} {\bf 386},
  \penalty0 1,  (2003).

\bibitem{Bogner:2009bt}
S.~K. Bogner, R.~J. Furnstahl, and A.~Schwenk, \emph{Prog. Part. Nucl. Phys.}
  {\bf 65}, \penalty0 94,  (2010).

\bibitem{Lepage:1997cs}
G.~P. Lepage.
\newblock \emph{{``How to Renormalize the Schr{\"o}dinger Equation'', Lectures
  given at 9th Jorge Andre Swieca Summer School: Particles and Fields, Sao
  Paulo, Brazil, February, nucl-th/9706029}},  (1997).

\bibitem{Bedaque:2002mn}
P.~F. Bedaque and U.~van Kolck, \emph{Ann. Rev. Nucl. Part. Sci.} {\bf 52},
  \penalty0 339,  (2002).

\bibitem{Bogner:2005sn}
S.~K. Bogner, A.~Schwenk, R.~J. Furnstahl, and A.~Nogga, \emph{Nucl. Phys. A}.
  {\bf 763}, \penalty0 59,  (2005).

\bibitem{Ramanan:2007bb}
S.~Ramanan, S.~K. Bogner, and R.~J. Furnstahl, \emph{Nucl. Phys.} {\bf A797},
  \penalty0 81,  (2007).

\bibitem{Jurgenson:2009qs}
E.~D. Jurgenson, P.~Navratil, and R.~J. Furnstahl, \emph{Phys. Rev. Lett.} {\bf
  103}, \penalty0 082501,  (2009).

\bibitem{Anderson:2010aq}
E.~R. Anderson, S.~K. Bogner, R.~J. Furnstahl, and R.~J. Perry, \emph{Phys.
  Rev.} {\bf C82}, \penalty0 054001,  (2010).

\bibitem{Roth:2011ar}
R.~Roth, J.~Langhammer, A.~Calci, S.~Binder, and P.~Navratil, \emph{Phys. Rev.
  Lett.} {\bf 107}, \penalty0 072501,  (2011).

\bibitem{Jurgenson:2010wy}
E.~Jurgenson, P.~Navratil, and R.~Furnstahl, \emph{Phys.Rev.} {\bf C83},
  \penalty0 034301,  (2011).

\bibitem{bohr69a}
A.~Bohr and B.~R. Mottelson, \emph{Nuclear Structure, Vol. 1}. (Benjamin,
  1969).

\bibitem{duguet02a}
T.~Duguet, P.~Bonche, P.-H. Heenen, and J.~Meyer, \emph{Phys. Rev. C}. {\bf
  65}, \penalty0 014310,  (2002).

\bibitem{duguet02b}
T.~Duguet, P.~Bonche, P.-H. Heenen, and J.~Meyer, \emph{Phys. Rev. C}. {\bf
  65}, \penalty0 014311,  (2002).

\bibitem{bender03b}
M.~Bender, P.-H. Heenen, and P.-G. Reinhard, \emph{Rev. Mod. Phys.} {\bf 75},
  \penalty0 121,  (2003).

\bibitem{ring80a}
P.~Ring and P.~Schuck, \emph{The Nuclear Many-Body Problem}. (Springer-Verlag,
  New-York, 1980).

\bibitem{bender03a}
M.~Bender, H.~Flocard, and P.-H. Heenen, \emph{Phys. Rev. C}. {\bf 68},
  \penalty0 044321,  (2003).

\bibitem{Margueron:2007uf}
J.~Margueron, H.~Sagawa, and K.~Hagino, \emph{Phys. Rev.} {\bf C77}, \penalty0
  054309,  (2008).

\bibitem{Yamagami:2008ks}
M.~Yamagami, Y.~R. Shimizu, and T.~Nakatsukasa, \emph{Phys. Rev.} {\bf C80},
  \penalty0 064301,  (2009).

\bibitem{Chamel:2010rw}
N.~Chamel, \emph{Phys. Rev.} {\bf C82}, \penalty0 014313,  (2010).

\bibitem{barranco04a}
F.~Barranco, R.~Broglia, G.~Colo', E.~Vigezzi, and P.~Bortignon, \emph{Eur.
  Phys. J.} {\bf A21}, \penalty0 57,  (2004).

\bibitem{gori05}
G.~Gori, F.~Ramponi, F.~Barranco, P.-F. Bortignon, R.~A. Broglia, G.~Col{\`o},
  and E.~Vigezzi, \emph{Phys. Rev. C}. {\bf 72}, \penalty0 011302(R),  (2005).

\bibitem{Pastore:2008zi}
A.~Pastore, F.~Barranco, R.~Broglia, and E.~Vigezzi, \emph{Phys.Rev.} {\bf
  C78}, \penalty0 024315,  (2008).

\bibitem{Idini:2011zz}
A.~Idini, F.~Barranco, E.~Vigezzi, and R.~Broglia, \emph{J.Phys.Conf.Ser.} {\bf
  312}, \penalty0 092032.

\bibitem{duguet04a}
T.~Duguet, \emph{Phys.\ Rev.} {\bf C69}, \penalty0 054317,  (2004).

\bibitem{Duguet:2007be}
T.~Duguet and T.~Lesinski, \emph{Eur. Phys. J. ST}. {\bf 156}, \penalty0 207,
  (2008).

\bibitem{Lesinski:2008cd}
T.~Lesinski, T.~Duguet, K.~Bennaceur, and J.~Meyer, \emph{Eur. Phys. J.} {\bf
  A40}, \penalty0 121,  (2009).

\bibitem{Hebeler:2009dy}
K.~Hebeler, T.~Duguet, T.~Lesinski, and A.~Schwenk, \emph{Phys. Rev.} {\bf
  C80}, \penalty0 044321,  (2009).

\bibitem{Baroni:2009eh}
S.~Baroni, A.~O. Macchiavelli, and A.~Schwenk, \emph{Phys. Rev.} {\bf C81},
  \penalty0 064308,  (2010).

\bibitem{Duguet:2010qw}
T.~Duguet, T.~Lesinski, K.~Hebeler, and A.~Schwenk, \emph{Mod. Phys. Lett.}
  {\bf A25}, \penalty0 1989,  (2010).

\bibitem{Lesinski:2011rn}
T.~Lesinski, K.~Hebeler, T.~Duguet, and A.~Schwenk, \emph{J.Phys.G}. {\bf G39},
  \penalty0 015108,  (2012).

\bibitem{entem03}
D.~R. Entem and R.~Machleidt, \emph{Phys. Rev. C}. {\bf 68}, \penalty0 041001,
  (2003).

\bibitem{bogner07a}
S.~K. Bogner, R.~J. Furnstahl, S.~Ramanan, and A.~Schwenk, \emph{Nucl. Phys.}
  {\bf A784}, \penalty0 79,  (2007).

\bibitem{Hebeler:2010xb}
K.~Hebeler, S.~Bogner, R.~Furnstahl, A.~Nogga, and A.~Schwenk, \emph{Phys.Rev.}
  {\bf C83}, \penalty0 031301,  (2011).

\bibitem{perezmartin08a}
S.~Perez-Martin and L.~M. Robledo, \emph{Phys. Rev. C}. {\bf 78}, \penalty0
  014304,  (2008).

\bibitem{soma11a}
V.~Som\`a, T.~Duguet, and C.~Barbieri, \emph{Phys. Rev.} {\bf C84}, \penalty0
  064317,  (2011).

\bibitem{soma11b}
V.~Som\`a, C.~Barbieri, and T.~Duguet, {Unpublished}.  (2011).

\bibitem{Nambu:1960tm}
Y.~Nambu, \emph{Phys. Rev.} {\bf 117}, \penalty0 648,  (1960).

\bibitem{Koltun:1972}
D.~S. Koltun, \emph{Phys. Rev. Lett.} {\bf 28}, \penalty0 182,  (1972).

\bibitem{baym62}
G.~Baym, \emph{Phys. Rev.} {\bf 127}, \penalty0 1391,  (1962).

\bibitem{soma11d}
V.~Som\`a, T.~Duguet, and C.~Barbieri, \emph{Journal of Physics: Conf. Ser.}
  {\bf 321}, \penalty0 01239,  (2011).

\end{thebibliography}

%\printindex[aindx]                 % to print author index
\printindex                         % to print subject index
\end{document}